**Self-passivation reduces the Fermi level pinning in the metal-semiconductor contacts**

Ziying Xiang[1,2], Jun-Wei Luo[1,2,*], and Shu-Shen Li[1,2]

[1]*State Key Laboratory of Semiconductor Physics and Chip Technologies, Institute of Semiconductors, Chinese Academy of Sciences, Beijing 100083, China*

[2]*Center of Materials Science and Optoelectronics Engineering, University of Chinese Academy of Sciences, Beijing 100049, China*

[*]Email: jwluo@semi.ac.cn.

**Abstract**

The metal-induced gap states (MIGS) are commonly believed to cause the strong Femi level pinning (FLP) in the metal-semiconductors contacts. Here, we unravel unambiguously that the dangling bonds-induced interface states play a crucial role, even comparable with MIGS. The first-principles calculations show that metal-Ge and metal-Si contacts should possess a similar FLP strength if they adopt an identical interface bonding configuration: the reconstructed bonding configuration renders Si and Ge having pinning factors of 0.16 and 0.11, respectively, and the ideal non-reconstructed bonding configuration gives them pinning factors of 0.05 and 0, respectively. We illustrate that Si favors the reconstructed bonding configuration, and Ge favors the ideal non-reconstructed bonding configuration after metal deposition. The self-passivation of the dangling bonds substantially reduces the interface gap states to give a much weaker FLP in the metal-Si contacts than in the metal-Ge contacts. We also demonstrate that the full passivation of the interface dangling bonds can further increase the pinning factor to 0.5 by further reducing the interface gap states. These findings shed new light on alleviating the Femi level pinning to lower the contact resistance for Si and emerging materials towards advanced semiconductor technology.

With the continued scaling of Si complementary metal-oxide-semiconductor (CMOS) technology, the influence of source/drain contact resistance on the performance of metal-oxide-semiconductor field-effect transistors (MOSFETs) has been increasingly significant due to a smaller contact area and is now becoming a major bottleneck for advanced transistor performance scaling [1,2]. The International Roadmap for Device and Systems (IRDS 2022) [3] sets a target for Si contact resistance below $10^{-9}$ Ω-cm$^2$, a goal that is challenging yet attainable. The challenge is primarily due to the strong Fermi level pinning (FLP) effect that causes a finite Schottky barrier height (SBH) $\Phi_{B,n}$ to form at the metal-Si interface. Meanwhile, the continued dimensional scaling of Si transistors also substantially degrades the channel mobility, which poses another great challenge and underscores the urgency to explore high-mobility channel materials, such as Ge [4], to sustain Moore's Law. Although Ge possesses superior carrier mobility as it has the highest hole mobility among all semiconductors, with electron mobility approximately three times higher than in Si [4], the extremely strong FLP effect at the metal-Ge interfaces [5] prevents it from replacing Si for developing next-generation low-power transistors. Two-dimensional (2D) semiconductors as channel materials could enable transistors with sub-10 nm gate length while maintaining sufficiently small subthreshold swing and leakage current with ultimate body thickness down to one atomic layer but also suffer from poor source and drain contacts with a considerably large Schottky barrier due to the Fermi level pinning (FLP) effect [6-9]. The strength of the FLP for a given semiconductor is usually characterized by a pinning factor $S$ [10]:

$$S = \frac{d\Phi_{B,n}}{d\phi_M},\qquad(1)$$

where $\phi_M$ is the work function (WF) of metals in contact. Experiments show that n-type Si has a pinning factor of $S = 0.16$ [11,12], whereas n-type Ge possesses $S = 0.02$ [5,13-15] the strongest FLP among semiconductors, manifesting the Fermi level firmly pinned at about 0.05 eV above the valence band maximum (VBM) and yielding a large constant n-type SBH (0.57±0.07 eV) for the Ge-metal interface irrespective of the metal in contact [16]. Given the striking similarities in both crystal structure and electronic properties between Si and Ge, it is particularly interesting to note that Ge has much stronger pinning than Si, which makes Ge unlikely to replace Si for an NMOS channel [5,17].

The FLP effect is well recognized as arising from an interface dipole caused by charge redistribution at the metal-semiconductor (MS) interface due to the metal-induced gap states (MIGS), which originate from the penetration of metal Bloch wave functions into the semiconductor band gap [18,19]. The occupation of the MIGS by electrons from the metal generates charges on the semiconductor surface, accompanied by induced opposing charges with the same amount on the metal surface. This charge redistribution establishes interface dipole across the MS interface, resulting in a deviation of the metal Fermi level from its anticipated position as predicted by the Schottky-Mott rule [20,21] and the pinning of the Fermi level towards a specific energy level, which is usually named the charge-neutrality level (CNL) [19,22,23]. Based on the electrical double-layer model, the pinning factor $S$ can be expressed as follows:

$$S = \left(1 + \frac{e^2 \delta_{it} D_{it}}{\varepsilon_{it}}\right)^{-1} \quad (2)$$

where $D_{it}$, $\delta_{it}$, and $\varepsilon_{it}$ are the density of interface gap states per unit area, effective distance away from the interface (associated with the decay length of interface states), and effective dielectric constant at the interface region, respectively [24]. It is straightforward to learn that a higher density of interface gap states $D_{it}$ gives rise to a smaller S and thus a stronger FLP effect [24]. The MIGS model also relates the $D_{it}$ and $\delta_{it}$ to the magnitude of band gap $E_g$ for a semiconductor [11,25], suggesting that Ge with a narrower band gap should exhibit a stronger FLP effect than Si due to higher density and larger decay length of MIGS [11]. The alleviation of the strong FLP by inserting an ultrathin dielectric insulator layer into the interface between metal and semiconductor (Si and Ge) has further proved the MIGS model, regarding that it can reduce the penetration of MIGS from the metal into the semiconductor [13,15,16,26]. However, the MIGS model has recently been questioned by both theoretical and experimental observations that show interface structure can strongly modify the SBH and FLP [27,28], which are unlikely to be explained in the context of the MIGS model since it regards FLP as purely an intrinsic property of the bulk semiconductors as it relies solely on bulk band gap or dielectric constant without considering the role of the specific structure of the interface. Therefore, the microscopic mechanisms underlying the strong FLP in the metal-semiconductor interfaces remain ambiguous. It prompts us to reconsider whether the difference in the FLP strength between Ge and Si originates from their difference in interfacial bonding structures instead of intrinsic MIGS to explore more feasible ways to overcome the contact resistance challenge for advanced semiconductor technology nodes.

In this work, by performing the first-principles calculations, we show that both Si and Ge have a similar FLP strength if they adopt the same interfacial bonding structure in contact with metals. Specifically, if a $p(2\times2)$ dimer reconstruction interfacial bonding configuration is adopted, the predicted pining factor $S = 0.16$ is in excellent agreement with the well-established experimental data of $S = 0.16$ for n-type metal-Si (001) interfaces [11,12]. However, this interfacial bonding configuration also results in metal-Ge (001) interfaces having a pinning factor $S = 0.11$ that is much larger than the experimental data $S = 0.02$ [5,13]. Interestingly, the ideal $c(1\times1)$ non-reconstructed interfacial bonding structure substantially reduces the pinning factor of Ge to $S = 0$ to reproduce the experimental data well, but it gives rise to the pinning factor of Si dropping to $S = 0.05$. We show that the self-passivation of the interfacial Si dangling bonds renders the $p(2\times2)$ dimer reconstruction to have a smaller density of interface states and, thus, a larger $S$ and a weaker FLP in comparison with the $c(1\times1)$ non-reconstructed interfacial bonding structure. We further illustrate that metal-Si (001) interfaces prefer the $p(2\times2)$ dimer reconstruction as expected [29,30], but the metal-Ge (001) interfaces prefer the $c(1\times1)$ non-reconstructed structure because it gains much less energy from the dimer reconstruction due to a longer bond length. We also demonstrate that we can substantially alleviate the FLP by increasing the pinning factor $S$ to near 0.5 by fully passivating the interface dangling bonds with, e.g., hydrogen atoms for both Si and Ge. Note that the residual effect hindering the approaching to the Schottky-Mott limit $S = 1$ should arise from the MIGS, which is commonly considered as the major factor for FLP. These findings shed new light on overcoming the contact resistance for advanced technology nodes and new materials for future transistors.

**Results**

**Schottky barrier height comparison of metal-Si and metal-Ge**

It is well documented that the topmost-layer atoms of the clean Si (001) and Ge (001) surfaces will undergo pairing spontaneously, yielding reconstructions in $c(2\times1)$, $c(4\times2)$, $p(2\times2)$, or $p(4\times1)$ configurations [30,31]. The $p(2\times2)$ dimer reconstruction has the lowest energy at low temperatures [32] and arises from the non-reconstructed ideal $c(1\times1)$ surface by pairing the topmost-layer atoms along the [110] direction to form alternating buckled asymmetric dimers: one atom of the dimers displaces away from the surface (dimer-up atom), while the other moves inward (dimer-down atom) [33,34]. Therefore, we calculate the SBHs of the metal-Si and metal-Ge interfaces by considering metal deposition on the CMOS-compatible Si (001) and Ge (001) substrates, which are both in either $p(2\times2)$ dimer reconstruction or ideal non-reconstructed $c(1\times1)$ surface, using the first-principles density functional theory (DFT) with Heyd-Scuseria-Ernzerhof hybrid functional (HSE06) approach [35,36] as implemented in the PWmat code [37]. For more details, see the supplementary information.

Figure 1 shows the first-principles calculation results of the SBHs at the metal-Si (001) and metal-Ge (001) interfaces against the WF of metals in contact for both considered surface configurations, in comparison with the well-established experimental data [5,11-14,38-44]. For the $p(2\times2)$ dimer reconstruction, one can see from Fig. 1(a) that the first-principles calculated SBHs of the metal-Si (001) interfaces are in reasonable agreement with the experimental measurements considering experimental data exhibiting a large scatter, but it is not true for the metal-Ge (001) interfaces. By fitting the theoretically predicted SBHs as a linear function of metal WF, we obtain a pinning factor $S = 0.16$ same as the experimentally reported value $S = 0.16$ for the metal-Si (001) interfaces [11,12]. The only exception is a slightly vertical shift (<0.1eV) in energy between two fitted curves, which may arise from the first-principles calculation uncertainties [45-47], e.g., the difference of the predicted electron affinity of bulk semiconductors from the experimental data. Fortunately, these uncertainties do not affect our conclusions, which are based on the slope of the SBHs (see supplementary Table SM-II for detailed SBHs). However, for the metal-Ge (001) interfaces with the same $p(2\times2)$ dimer reconstruction, the predicted pinning factor $S = 0.11$ is unexpectedly large, which is only 0.05 smaller than that of the metal-Si interfaces and thus is too large to account for the extremely strong FLP ($S = 0.02$) observed experimentally in the metal-Ge contacts [5,13,14], as shown in Fig. 1(c). Interestingly, as we change the interface configuration from the $p(2\times2)$ dimer reconstruction to the ideal $c(1\times1)$ non-reconstructed structure, the first-principles calculations reproduce the experimentally measured pinning factor of the metal-Ge contacts by substantially reducing the predicted pinning factor from $S = 0.11$ to $S = 0$ (Fig. 1(d)). Surprisingly, the ideal $c(1\times1)$ non-reconstructed structure also remarkably reduces the pinning factor of the metal-Si interfaces to $S = 0.05$, manifesting that such interface configuration artificially makes Si have an extremely strong FLP comparable to Ge.

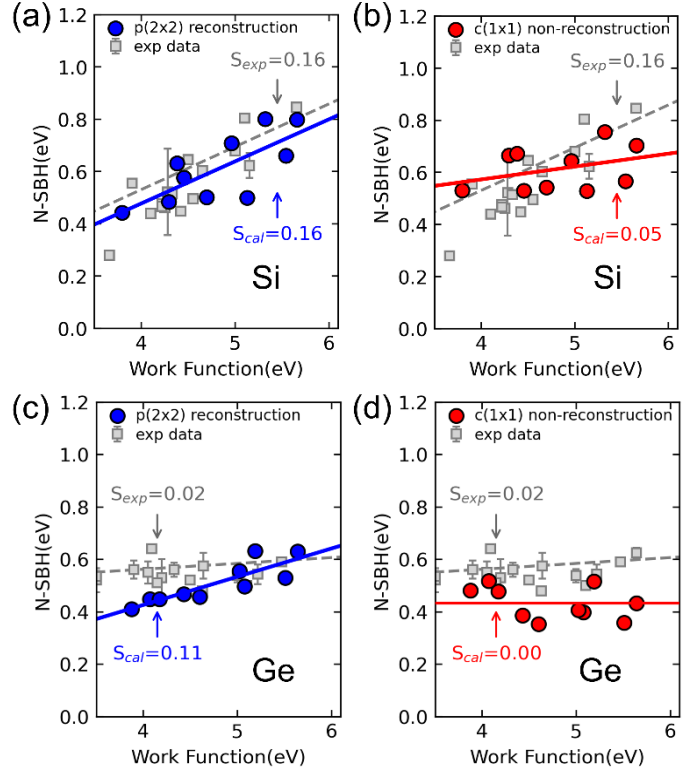

**Fig. 1**. First-principles method calculated n-type Schottky barrier height (N-SBH) of metal-semiconductor contacts with non-reconstructed and reconstruction interface configurations. **a, b** Metal-Si (001) interfaces in (a) $p(2\times2)$ dimer reconstruction interface configuration, and (b) $c(1\times1)$ non-reconstructed interface configuration. **c, d** Metal-Ge (001) interfaces in (c) $p(2\times2)$ dimer reconstruction interface configuration, and (d) $c(1\times1)$ non-reconstructed interface configuration. Grey squares with error bars represent the experimental data.

Consequently, we show that the interfacial atomic bonding configuration plays a crucial role in determining the FLP strength since metal-Ge and metal-Si interfaces have similar FLP strength if they adopt an identical interface configuration. It implies that the metal-Si contacts prefer a reconstruction interface with a weaker FLP ($S = 0.16$) [11,12], and the metal-Ge contacts prefer an ideal non-reconstructed interface with an extremely strong FLP ($S = 0.02$) [5,13,14]. This finding is strikingly in contrast with the common wisdom according to the MIGS models [13,18,19], which suggest the FLP strength to be an intrinsic property of bulk semiconductors without the need to consider the detailed atom arrangements at the interface [13,19]. Evidence has been accumulated that metal deposition may destroy the reconstructed surface structure of the substrate [48,49], although both clean Si (001) and Ge (001) surfaces are predicted theoretically [30,32] and observed experimentally [50] to be stabilized in the $p(2\times2)$ dimer reconstruction [32]. For instance, experiments have shown that Ba adsorption on the (001) Ge surface results in a change from the reconstruction surface to the $c(1\times1)$ non-reconstructed surface [51]. It is worth noting that the small discrepancy between our first-principles calculations and experimental data may arise from the admixture of the $c(1\times1)$ non-reconstructed configuration and dimer reconstruction configuration in reality as well as the limitations of the first-principles calculations.

**The role of two types of interface gap states in the FLP effect**

We first attempt to reveal the microscopic mechanism responsible for the effect of interfacial atomic bonding configurations on the FLP strength before addressing why metal-Si contacts prefer a reconstruction interface but metal-Ge contacts prefer a non-reconstructed interface. To do so, we take Ag as a prototypical metal using an Ag-Ge supercell containing 9 Ag atomic layers and 13 Ge atomic layers in the [001]-direction, as shown in supplementary Fig. SM-2, to examine the metal-Ge (001) interface band structure by projecting the electronic states of the Ag-Ge (001) interface on the topmost Ge layer. Notably, its projected density of states (PDOS) is more critical since the density of interface gap states $D_{it}$ governs the pinning factor $S$ according to equation (2). Figure 2(a) shows that in the $c(1\times1)$ non-reconstructed configuration Dirac-cone-like bands cross at the X-point, which, however, is absent in the $p(2\times2)$ dimer reconstruction configuration (Fig. 2(b)). It causes the total density of the interface states near the Fermi level to get reduced by changing the arrangement of the interfacial Ge atoms from the $c(1\times1)$ non-reconstructed configuration to the $p(2\times2)$ dimer reconstruction. It should account for a larger pinning factor $S$ and weaker FLP strength observed in the $p(2\times2)$ dimer reconstruction interface configuration.

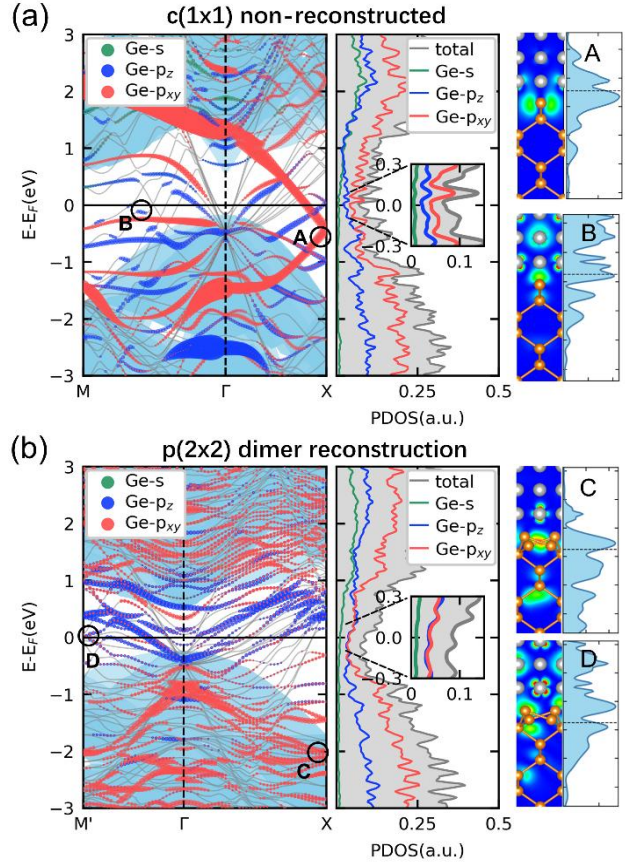

**Fig. 2.** The first-principles calculated interface band structure of the metal-Ge (001) interface. **a, b** The Ag-Ge (001) interface electronic states are projected onto the first layer of Ge atoms (left panel) and the projected density of states (PDOS) of all Ge layers (middle panel) for (a) $c(1\times1)$ non-reconstructed interface configuration and (b) $p(2\times2)$ dimer reconstruction interface configuration, respectively. In the left panels, the light-blue shadow represents the bulk band structure projected to the (001) surface BZ; the gray lines denote the interface band structure, and the colored dots represent Ge atomic orbitals with dot size accounting for the relative amplitude. In the middle panels, the insets zoom the PDOS near the Fermi level. The right panels show the partial charge distribution for typical interface states marked by black circles in the left panels. The orange spheres denote Ge atoms, and the silver spheres denote Ag atoms.

To better understand the FLP mechanism, it is necessary to identify these gap states. From the partial charge distribution of these interface gap states in real space, we can distinguish them either as metal-induced gap states (MIGS) [18,19] or dangling-bond-induced surface states (DBSS) [52].

Specifically, MIGS originates from the penetration of metal Bloch wave functions into the semiconductor band gap [18,19], resulting in charges distributing across the entire metal region and gradually decaying towards the semiconductor side [10]. Whereas DBSS emerges from the surface dangling bonds owing to the crystal lattice termination at the surface, and thus, its charges localize predominantly in the interface region and decay towards both sides [53]. Subsequently, we can identify Dirac-cone-like bands crossing at the X-point in the $c(1\times1)$ non-reconstructed configuration as DBSS (see the right panel in Fig. 2(a)) and derived from the rehybridized $p_{xy}$ states of surface Ge atoms (see the left panel in Fig. 2(a) and will be discussed below) [54,55]. We can further ensure the identification of the DBSS in Fig. 2(a) by referring to the surface states of the bare Ge (001) surface (supplementary Fig. SM-6), which possesses nearly identical Dirac-cone-like bands in both the energy dispersion and partial charge distributions. It confirms Dirac-cone-like bands arising from the $p_{xy}$-states of the surface dangling bonds induced by the semiconductor surface termination.

We also identify some interface gap states, e.g., indicated by B in Fig. 2(a), arising from MIGS as expected. Fig. 3 schematically shows that when changing from the $c(1\times1)$ non-reconstructed configuration to the $p(2\times2)$ dimer reconstruction configuration, the in-plane dimerization of the Ge surface dangling bonds will shift most $p_{xy}$ orbital states out of the bandgap to below the VBM or above the CBM, thereby preventing the formation of Dirac-cone-like bands, as shown in Fig. 2(b). Specifically, these rehybridized $p_{xy}$ states (such as indicated by C in Fig. 2(b)) are predominantly localized within the dimer region and resonate with the Ge bulk states, yielding the surface resonance states [52,53]. On the other hand, the $p_z$-MIGS orbital states do not exhibit significant energy shifts and remain inside the band gap as changing the interface configurations (such as indicated by B and D). Although both DBSS and MIGS contribute to the total density of the interface gap states $D_{it}$ for FLP effect, the contribution of the MIGS is unlikely to be changed by altering the interface structures due to it is an intrinsic property of bulk semiconductors. It implies that the engineering of these DBSSs, giving rise to the Dirac-cone-like bands composed of $p_{xy}$ states, are mainly responsible for the reduction of total density of interface gap states $D_{it}$ near the Fermi level shown in Fig. 2, and thus a smaller interface dipole potential and a weaker FLP strength by changing the $c(1\times1)$ non-reconstructed configuration to the $p(2\times2)$ dimer reconstruction interface configuration. This variation is also observed in other Ge-metal interface and Si-metal interface (see

Fig. SM-4), with moderate differences arising from perturbations induced by different metals. Nonetheless, the overall trend of a decrease in the $p_{xy}$ DOS, attributed to the disappearance of the Dirac-cone-like band due to in-plane dimer-reconstruction, remains a general phenomenon.

**Microscopic mechanism governing interface states evolution**

We now turn to explain the formation and change of DBSS in terms of atomic orbitals, using schematic diagrams as shown in Fig. 3. The ideal $c(1\times1)$ non-reconstructed (001) surface results from cleaving along the (001) plane of the cubic unit cell of the diamond structure, leaving two dangling bonds for each surface atom and one valence electron in each dangling bond. For each surface atom, two $sp^3$ hybrid orbitals from two dangling bonds will rehybridize into one $p_{xy}$ and one $sp_z$ states. The energy level of the rehybridized $p_{xy}$ state rises from the $sp^3$ level to the atomic level of the outermost $p$-orbitals (3$p$ in Si and 4$p$ in Ge), and the rehybridized $sp_z$ state goes down to the average of the outermost $s$- and $p$-orbital atomic levels, respectively. Then, the two valence electrons owned by the two dangling bonds in each surface atom transfer to the lower-lying rehybridized $sp_z$ state to make it fully occupied and leave the higher-lying rehybridized $p_{xy}$ state completely empty, as shown in Fig. 3(a). The higher-lying rehybridized $p_{xy}$ states from two adjacent surface atoms will couple together to give rise to the bonding $\sigma$ and antibonding $\sigma^*$ states by shifting upward and downward in their energy levels. The periodic potential perturbation within the (001) plane further broadens these discrete $\sigma_{p_{xy}}$ and $\sigma^*_{p_{xy}}$ energy levels into Dirac-cone-like bands for the bare Ge (001) surface (and similarly for the bare Si (001) surface, as shown in Fig. SM-5 and Fig. SM-6) [54]. In contrast, Fig. SM-5 shows that the energy band of the rehybridized $p_z$ surface states exhibit negligible dispersion due to its much weaker coupling between adjacent $p_z$ orbitals within the (001) plane.

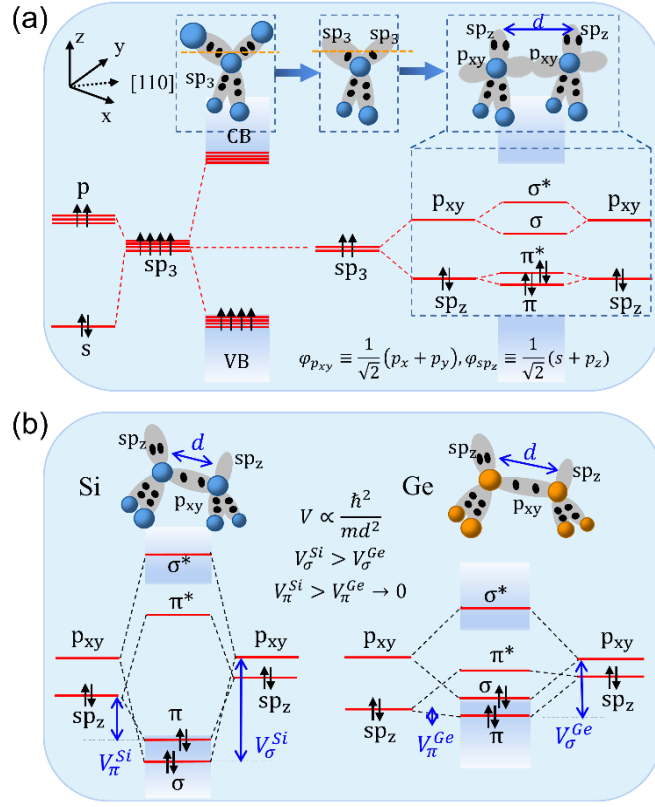

**Fig 3. Schematic diagram illustrating the coupling and the corresponding energy level shift of the dangling bonds induced surface states.** **(a)** The process of cleave along the (001) plane in bulk Si or Ge, and the rehybridization of the $sp_3$ states of the dangling bonds into $p_{xy}$ and $sp_z$ states, which will couple together to form new dangling-bonds-induced surface states. $\varphi_{pxy}$ and $\varphi_{sp_z}$ represent the rehybridized wave functions from original $sp_3$ orbitals. **(b)** The coupling of adjacent (001) surface atomic orbitals in the [110] direction resulting from the dimerization in the $p(2\times2)$ dimer reconstruction configuration on the Si (001) and Ge (001) bares surfaces before metal deposition on it. Note an asymmetric dimer structure causes the rehybridized $sp_z$ and $p_z$ states of the dimer-down shift up relative to that of the dimer-up [29]. Here, $d$ and $V$ represents the distance of two dimer atoms and the energy level shift caused by various types of orbital-coupling.

When two adjacent surface atoms come closer to each other along the [110] direction to form a dimer as a result of the transition from the $c(1\times1)$ non-reconstructed structure to the $p(2\times2)$ dimer reconstruction, as shown in Fig. 3(b), their rehybridized $p_{xy}$ states significantly overlap end to end and develop into the $\sigma$ bonding and $\sigma^*$ antibonding states, opening energy gap at the X-point for the Dirac-cone-like bands and moving from inside the band gap to below the VBM or above the CBM of the bulk semiconductor, respectively. At the same time, their low-lying rehybridized $sp_z$ states overlap to form $\pi$ bonding and $\pi^*$ antibonding states but with less energy shift. Note that the surface dimers would further tilt to have an asymmetric dimer structure with one atom up (named dimer-up) and another atom down (named dimer-down), rendering the rehybridized $sp_z$ state of the dimer-down atom to have more $p_z$ character [29]. The valence electrons occupied the lower-lying rehybridized $sp_z$ state in the $c(1\times1)$ non-reconstructed structure transfer to their even lower $\sigma$ and $\pi$ bonding states, which further lowers the surface energy of the $p(2\times2)$ dimer reconstruction and thus explains why both clean Si (001) and Ge (001) surfaces stabilized in the $p(2\times2)$ reconstruction at T = 0 K [32]. More details are given in supplementary Fig. SM-5. The energy shift caused by the dimerization of surface dangling bonds reduces the interface gap states to alleviate the FLP effect in the $p(2\times2)$ dimer reconstruction configuration, yielding a self-passivation effect of the surface dangling bonds.

The energy gain from the $p(2\times2)$ dimer reconstruction on the (001) surface is governed by the energy level separation between the bonding and antibonding states of the dimer, which is inversely proportional to the square of the dimer bond length [56] or the lattice constant of bulk semiconductors. Because the lattice constant of Ge is 4.3% larger than that of Si, energy gain from the dimer reconstruction on the Ge (001) surface is less than that on the Si (001) surface, as shown schematically in Fig. 3(b). This argument is supported by the reduced surface reconstruction energy, which shows that the dimer formation energy is 1.24 eV/dimer for Ge and 1.77 eV/dimer for Si according to the first-principles calculations. Thus, the $p(2\times2)$ dimer reconstruction is expected to be less stable in Ge than in Si, consistent with previous studies [29,57,58]. Specifically, weaker dimer bonding on the reconstructed Ge (001) surface has been confirmed by measurements of the dimer bond length and bonding strength [29], manifesting that dimer bonds in Ge exhibit nearly single-bond character, while they in Si exhibit some double-bond character. Subsequently, on the Ge (001) surface, metal deposition is more likely to perturb the $\sigma$ bonding state, shifting it above the $\pi^*$ anti-bonding state (see supplementary Fig. SM-5(e)-(h)). This energy level reversal would result in no energy gain from dimerization, potentially causing the Ge (001) surface to revert to the ideal $c(1\times1)$ non-reconstructed configuration. To further explore this behavior, we perform molecular dynamics simulations of lattice vibrations at 300K. The result in supplementary Fig. SM-7 shows that the fluctuations in the dimer bond lengths at the $p(2\times2)$ dimer reconstruction Ag-Ge (001) interface gradually increase over time, indicating a progressive weakening of the dimer bonds, whereas it remains relatively stable over time in the Ag-Si (001) interface. Although the simulation timescale is limited to approximately 7000 fs (~7 ps) and only captures the initial stages of bond fluctuation after the structure allowed to relax at 300 K, the pronounced differences suggest that Ge dimers in the metal-Ge (001) interfaces are less stable compared to Si dimers in the metal-Si (001) interfaces. Therefore, both thermodynamic and kinetic results confirm that metal-Si interfaces favor the $p(2\times2)$ interface reconstruction, while the metal-Ge interfaces tend toward the $c(1\times1)$ non-reconstructed structure.

To generalize our conclusion, we also extend our investigation to diamond, another group IV semiconductor with an ultrawide bandgap of $Eg$ = 5.4 eV. Both theoretical calculations and experimental measurements show that large pinning factor of S~0.35 [59-62] arises from the self-passivation effect by the reconstruction in the diamond (001) surface, since diamond bond length is 34% shorter than Si bond length and thus the diamond (001) surface is better-preserved in dimer reconstruction after metal deposition [29,57]. Otherwise, our first-principles results show that the metal-diamond (001) interfaces have a much smaller pinning factor of S~0.14 using the $c(1\times1)$ non-reconstructed configuration (see Fig. SM-9 and calculation details in the Supplementary). We note that from Ge, Si, to C, the first-principles calculations predicted pinning factor S increases from 0.0, through 0.05 to 0.14 using the $c(1\times1)$ non-reconstructed configuration, and increases from 0.11, through 0.16 to 0.31 using the $p(2\times2)$ dimer reconstruction. Such increase is inversely proportional to the magnitude of bandgap, which is because the density of interface gap states $D_{it}$ given in Eq. (2) is inversely proportional to $Eg$.

**The full-passivation of the dangling bonds**

We have illustrated that the dimerization-induced self-passivation of the surface dangling bonds is a key factor in reducing the interface gap states to alleviate the FLP in metal-Si interfaces relative to metal-Ge interfaces. Considering the surface dangling bonds have not yet been fully passivated even in the $p(2\times2)$ interface reconstruction, as shown in Fig. 2(b), we postulate that we can further alleviate the FLP by improving the passivation of the surface dangling bonds using, e.g., hydrogen atoms, which have been successfully used to shift all surface states out from the band gap on bare Si and Ge surfaces [63-65]. To validate this postulation, we have terminated each surface dangling bond by a hydrogen atom using a bond length of 1.51 Å for the Si-H bond and 1.56 Å for the Ge-H bond, a configuration found to be stable by Northrup [66]. Figure 4(a) shows that the H-passivation remarkably increases the pinning factor $S$ from 0.16 to 0.5 for the metal-Si interfaces, accompanied by a significant reduction in the interface gap states, as shown in supplementary Fig. SM-8. This is also true for the metal-Ge interfaces as the pinning factor $S$ increased from 0.11 to 0.45, as shown in Fig. 4(b). Consequently, we have explicitly demonstrated the critical role of the surface dangling bonds-induced interface gap states on the FLP in MS interfaces. Considering the pinning factor increases up to 0.5 for the metal-Si interfaces by eliminating almost all the interface gap states arising from the dangling bonds, the residual effect hindering the approaching of the Schottky-Mott limit $S=1$ should arise from MIGS, which is commonly regarded as the prime factor for FLP.

Note that although $S = 0.5$ is still way from the Schottky-Mott limit $S =1$, it enables us to have a substantial SBH tunability to overcome the challenge of contact resistance in advanced technology nodes [2]. For instance, after full dangling bonds passivation, we can now use the low-work-function metals (e.g., Ti, Ta, Mg, Ag, Cu, as shown in Fig. 4) to achieve near-zero Schottky barrier heights for n-type Si and Ge, minimizing contact resistance. Optimizing both non-metallic atom passivation for DBSS elimination and compatible low-work-function metals is key for future high-performance, CMOS-compatible transistors.

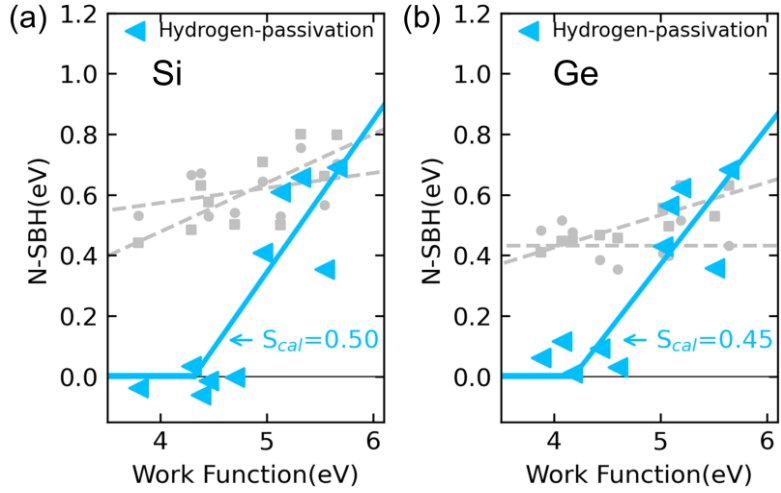

**Fig. 4.** First-principles predicted n-type Schottky barrier height (N-SBH) after the complete passivation of the surface dangling bonds by hydrogen atoms. **(a)** Metal-Si interfaces and **(b)** metal-Ge interfaces. For comparison, the grey dots and squares represent the predicted values in the ideal $c(1\times1)$ non-reconstructed configuration and the $p(2\times2)$ dimer reconstruction, respectively, as shown in Fig. 1.

**Discussion**

In addition to passivating the surface dangling bonds, there is another way to have the DBSS out from the middle bandgap to alleviate the FLP. Different from the covalent group IV semiconductors such as C, Si, and Ge, where a dangling bond state with an unpaired electron is formed near the middle of the band gap, as shown in the bottom panel of Fig. 5(b), the dangling bond states in ionic group II-VI semiconductors are near the CBM or VBM [53,67]. Because the CBM in strong ionic semiconductors is mainly made of the unoccupied orbitals of the cation and the VBM is mainly made of the occupied orbitals of the anion; therefore, the energy levels of DBSS, that are, either the non-bonding states of the cation or the non-bonding states of the anion, are formed near the CBM or VBM, allowing the dangling bonds to act as a shallow defects (the top panel of Fig. 5(b)). The density of DBSS in the middle bandgap is remarkably reduced as increasing the bandgap. Our calculated density of states for these dangling bonds (see Fig. 5(c)) in bare surfaces confirms this trend: for ZnO and SrO, DBSS is vanishing within the band gap, and for GaAs and BAs, the DBSS appeared but significantly reduced compared with Si and diamond. This reduction in DBSS for highly ionic materials directly correlates with the much larger pinning factor *S* observed for II–VI compounds, such as ZnO (*Eg* = 3.44 eV, *S*~0.57) and SrO (*Eg* = 5.79 eV, *S*~0.93) compared to covalent group IV covalent semiconductors, particularly diamond (*Eg* = 5.4 eV, *S*~0.35), despite their comparable Eg, as shown by the literature-reported *S* in Fig. 5(a) [11,42,59-62,68,69].

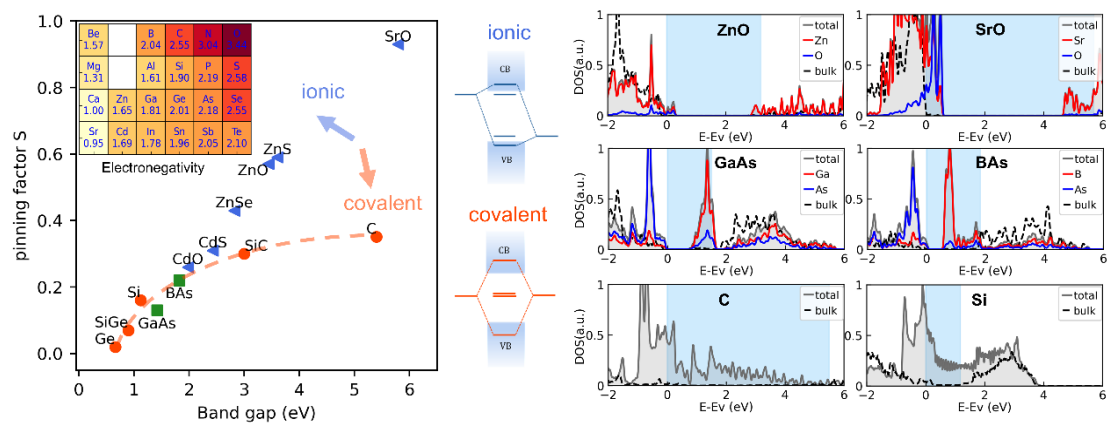

**Fig. 5. Pinning factor *S* and its relationship with the dangling bonds.** (**a**) The pinning factor *S* of various semiconductors, with an inset at the upper-left corner indicating the electronegativity values of different elements. (**b**) Schematic illustration of the bandgap formation mechanisms in covalent (bottom) and ionic (top) semiconductors and the typical energetic position of dangling-bonds-induced surface states, which is at the middle band gap in covalent semiconductor and near the CBM or VBM in the ionic semiconductor. (**c**) The density of states projected onto the surface layer for bare, unreconstructed, nonpolar surfaces of ZnO, SrO, GaAs, BAs, C and Si, where the x-axis denotes the energy relative to the valence band maximum *Ev*, and the light-blue shaded region indicates the band gap.

Thus, we have established a preliminary framework to generalize the role of DBSS in FLP across a boarder range of materials. Nevertheless, more in-depth studies are necessary to fully validate and extend this mechanism in future research. Notably, in highly ionic group II-VI semiconductors, the effect of the surface itself or the reconstruction process is relatively minor, as these are weak perturbations compared to the strong Coulomb forces. As a result, the reconstruction does not significantly influence the electronic structure [53,70]. In contrast, in more covalent semiconductors, the FLP is strongly affected not only by MIGS but also by DBSS, and thus the influence of self-passivation on the pinning strength is more pronounced as previously discussed. Overall, these findings point to a stepwise design principle for interface engineering to alleviate FLP by selectively modifying DBSS, including (1) self-passivation, such as in-plane dimerization, which shifts DBSS away from the mid-gap; (2) extrinsic full-passivation through strong bonding with non-metallic atoms, further reducing DBSS within the gap; and (3) fundamental suppression of DBSS formation by increasing the material's ionicity.

We note that both Nishimura et al. [13] and Kuzmin et al. [71] found that the modification of the atomic arrangement by altering the surface condition changes hardly the FLP strength or the pinning position and thus concluded that FLP of metal-Ge contacts is an intrinsic property and rises from MIGS. We must stress that their results are due to poor surface treatments, which may not effectively reduce the DBSS, considering Kuzmin's photoemission measurements showed little changes in the total surface component of core-level spectra by changing surface treatments [71]. However, it should also be emphasized that we do not disprove the significance of MIGS on the FLP in Ge and Si, although we attribute their different FLP to their different interface bonding configuration. The pinning factor $S$ is consistently predicted to be slightly larger (within 0.05) in Si than that in Ge for the same configurations, as shown in Fig. 1, indicating the effect of MIGS on FLP since Si has a larger band gap than Ge and, thus, a smaller density of MIGS. This is consistent with studies from Nishimura et al., where they find another effective method of using metals with low free-electron density such as Bi or germanide to substantially reduce MIGS thus alleviating FLP [14,15,72]. Besides, inserting an ultrathin insulator layer between the metal and the semiconductor has also been explored to alleviate the FLP by reducing the MIGS via suppressing the wave function tailing of the metal into the band gap of the semiconductor [17]. The ultrathin insulator insertion approach has been demonstrated to increase the pinning factor $S$ of the metal-Si and metal-Ge contacts up to 0.24 [73] and 0.3 [15,74], respectively. The residual FLP should be due to the effect of DBSS.

Since MIGS is an intrinsic property of semiconductors and its contribution to $S$ cannot be altered by hydrogen-passivation, we can roughly suppose that the increase of $S$ from near 0 to around 0.5 by hydrogen passivation for both Si and Ge is due to the elimination of DBSSs, while the remaining factor to hinder the approach of the Schottky-Mott limit $S = 1$ is due to residual MIGS.

This finding offers new ways to improve device performance, particularly for Ge transistors, by alleviating FLP strength. For instance, after modulating the interface atomic structure configuration, the large contribution of DBSS to FLP strength can be reduced without increasing the tunneling resistance and adding complexity to the next generation gate-all-around field effect transistor (GAAFET) technology; these two factors are usually suffered by the state-of-the-art approach of inserting an ultrathin insulator oxide layer into the metal-semiconductor interface to alleviate the FLP [15,75]. Crucially, the design principle governing hydrogen passivation, which achieves DBSS elimination through strong σ-bonding, also extends to other non-metallic atoms. Halogen (e.g., Cl) and chalcogen (e.g., S) atoms have also been reported to be capable of strongly binding to Si and Ge dangling bonds in experiment [76,77], and should also show notable passivation effectiveness. From a device integration perspective, at typical CMOS processing temperature (>400 °C), hydrogen passivation is thermally unstable on both Si and Ge [78,79], whereas Cl and S passivation provide much better stability and effective interface states suppression [76], which might be better choices for DBSS passivation. Therefore, balancing DBSS passivation, interface stability, and minimal impact on electrical transport remains a key challenge for future interface engineering studies.

In summary, we have revealed that the self-passivation of the interface dangling bonds is a primary factor in governing the different FLP strengths between the metal-Si and metal-Ge contacts since the former prefers the *p*(2×2) dimer reconstruction interface and the latter prefers the *c*(1×1) non-reconstructed interface structure. We have also demonstrated that utilizing hydrogen atoms to passivate the surface dangling bonds completely can further significantly increase the pinning factor *S* to 0.5, and the residual FLP should arise from the commonly suggested MIGS [11,13]. Thus, we shed new light on lowering the contact resistance for advanced technology nodes via modulating the FLP strength by passivating the interface bonding bonds without the need for an ultrathin dielectric layer [17,39], thereby avoiding the associated increase in tunnel resistance [15,75].

**Data availability**

The results data used for plotting the figures in this study are included in the Supplementary Information and have been deposited in the Figshare database. The raw DFT data generated in this study are available upon request from the corresponding author due to their large size and restricted access.

**Author Contributions**

J.W.L and S.S.L conceived the research concept and supervised the project. Z.Y.X performed the calculations and analyzed the results. Z.Y.X and J.W.L wrote the manuscript.

**Acknowledgements**

This work was supported by the National Natural Science Foundation of China (NSFC) under Grant Nos. 11925407 and the CAS Project for Young Scientists in Basic Research under Grant No. YSBR-026.

**Competing Interests**

The authors declare no competing interests.

**Self-passivation reduces the Fermi level pinning in the metal-semiconductor contacts**

**SUPPLEMENTAL MATERIAL**

Ziying Xiang[1,2], Jun-Wei Luo[1,2,*], and Shu-Shen Li[1,2]

[1]*State Key Laboratory of Semiconductor Physics and Chip Technologies, Institute of Semiconductors, Chinese Academy of Sciences, Beijing 100083, China*

[2]*Center of Materials Science and Optoelectronics Engineering, University of Chinese Academy of Sciences, Beijing 100049, China*

*Email: jwluo@semi.ac.cn.

**Modeling and computation details:**

Metal-Si (metal-Ge) interfaces in the (001) orientation consist of at least 13 atomic layers of Si (Ge) and 9 layers of metals. We selected 10 metals including Ag, Au, Pt, Ta, Mg, Cu, Ir, Pd, Rh, and Ti. Our primary goal is to investigate the trends in the Schottky barrier height (SBH) and Fermi level pinning (FLP) strength variations, which are primarily governed by the metal work function (WF) rather than specific metal structure arrangements. To maintain consistency, metals with face-centered cubic structures were selected despite potential discrepancies in experimental phase structures. We kept the lateral lattice parameters of Si (Ge) slabs fixed to their calculated equilibrium bulk constants, and we rotated all the metal lattices by 45° and applied lateral strain to match the lattice parameters of Si (Ge) slabs. This approach ensures uniformity in interface atomic structures across all interfaces. Given the alternations in metal lattice parameters, we recalculated the corresponding new WF of the metals (similar calculations can also be found in the previous studies [1,2]). Although the calculated WFs differ from experimental values, this does not affect our analysis, as our primary focus is on the pinning factor $S$, which is determined by the rate of change of SBH with respect to WF, rather than the WF of an individual metal. Further calculation details are provided in Table SM-I and Table SM-II.

Figure SM-1 presents the optimized surface structures of three configurations (only the Ge surfaces are shown, as the Si surfaces are similar). For both the $c(1\times1)$ non-reconstructed configuration and the $p(2\times2)$ dimer reconstruction configuration, we tested the preferred bonding configuration, which was found to be at the hollow site. For the hydrogen-passivated structure, we selected the 2×1 monohydride phase, known for its stability in hydrogen adsorption [3,4]. The preferred bonding configuration shifted to the top site. All surface structures were optimized using PBE functionals. We also optimized the displacements between the metal and semiconductor slabs to minimize the total energy. To better understand the correlation between the FLP effect and the ideal interface bonding structure, we fixed all the atomic positions. This strategy allowed us to control additional variables that typically arise during atomic relaxation, thereby avoiding the complexity, diversity, and non-repeatability of real interface structures. For instance, in actual experiments, varying preparation temperatures can alter the interface atomic arrangement, leading to changes that are non-repeatable. Additionally, relaxation can significantly modify metal surface structures, potentially altering the originally calculated WF and leading to discrepancies with the calculated SBH. By isolating these factors, we could more clearly identify the impact of the interface bonding structure on the SBH and FLP.

In bulk calculation with HSE functional [5,6], we performed band structure calculation with dense reciprocal space grids of 10×10×10 and adjusted the mixing coefficient a to obtain the correct experimental band gap (For Si, when a=0.25, $E_g$=1.17eV) and for Ge, when a=0.18, $E_g$=0.67eV). In the supercell calculations, considering the computational cost of HSE functional calculation, we used reciprocal space grids of 4×4×1 in the self-consistent calculation, and 10×10×1 in the following non-self-consistent calculations.

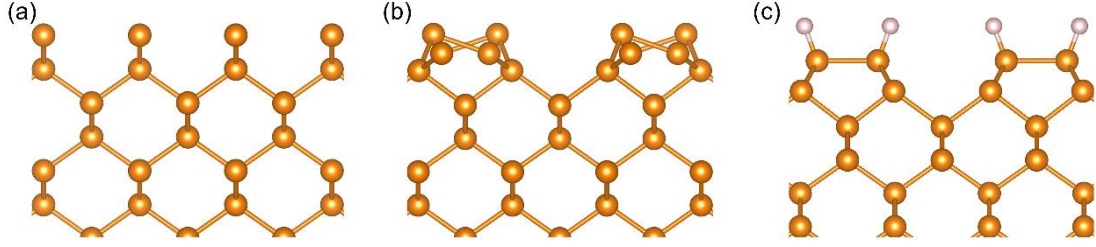

Fig. SM-1. Schematic atomic structures of Ge (001) surface with (a) $c(1\times1)$ non-reconstructed configuration, (b) $p(2\times2)$ dimer reconstruction configuration and (c) hydrogen-passivated configuration. The orange and pink spheres represent Ge and H atoms, respectively.

Table SM-I. The structure and the corrected metal work functions. Calculated metal equilibrium lattice parameters $a_M$, the lattice mismatch between the metals and Si (Ge), the experimental work function (exp WF) [7,8], and the calculated work function (cal WF). Each metal lattice is rotated by 45° to better match the substrate lattice. The negative values of lattice mismatch indicate that the rotated metals have smaller lattice parameters compared to Si (Ge). The cal WFs are adjusted with the lattice parameters equal to the lattice parameter of Si ($a_{Si}$ = 5.44Å) and Ge ($a_{Ge}$ = 5.62Å), respectively. The exp WFs listed in the last column are not all obtained from the crystalline samples in the (001) orientation, with some being from polycrystalline samples.

| Metal | $a_M$ (Å) | Si | | Ge | | exp WF (eV) |
| --- | --- | --- | --- | --- | --- | --- |
| | | lattice mismatch (%) | cal WF (eV) | lattice mismatch (%) | cal WF (eV) | |
| Ag | 4.14 | 7.63 | 4.29 | 4.16 | 4.07 | 4.26 |
| Au | 4.15 | 7.8 | 5.12 | 4.33 | 5.08 | 5.10 |
| Pt | 3.94 | 2.27 | 5.66 | -1.03 | 5.64 | 5.65 |
| Ta | 4.21 | 9.47 | 4.7 | 5.95 | 4.6 | 4.25 |
| Mg | 4.51 | 17.13 | 3.8 | 13.36 | 3.88 | 3.66 |
| Cu | 3.57 | -7.27 | 4.46 | -10.25 | 4.43 | 4.65 |
| Ir | 3.83 | -0.51 | 5.54 | -3.71 | 5.51 | 5.27 |
| Pd | 3.93 | 2 | 5.32 | -1.28 | 5.19 | 5.12 |
| Rh | 3.79 | -1.51 | 4.96 | -4.68 | 5.02 | 4.98 |
| Ti | 4.1 | 6.59 | 4.38 | 3.15 | 4.17 | 4.33 |

Table SM-II. Calculated n-type Schottky barrier height (N-SBH) data of 10 metal-Si (Ge) (001) contacts in three different interface configurations.

| Metal | N-SBH (eV) | | | | | |
|---|---|---|---|---|---|---|
| | Si | | | Ge | | |
| | $c(1\times1)$ | $p(2\times2)$ | H | $c(1\times1)$ | $p(2\times2)$ | H |
| Ag | 0.67 | 0.49 | 0.03 | 0.51 | 0.42 | 0.11 |
| Au | 0.53 | 0.50 | 0.61 | 0.40 | 0.49 | 0.56 |
| Pt | 0.70 | 0.80 | 0.69 | 0.43 | 0.63 | 0.68 |
| Ta | 0.56 | 0.57 | 0 | 0.37 | 0.51 | 0.03 |
| Mg | 0.54 | 0.46 | -0.04 | 0.49 | 0.44 | 0.06 |
| Cu | 0.52 | 0.55 | -0.01 | 0.37 | 0.43 | 0.09 |
| Ir | 0.58 | 0.71 | 0.35 | 0.37 | 0.57 | 0.36 |
| Pd | 0.77 | 0.83 | 0.66 | 0.52 | 0.64 | 0.62 |
| Rh | 0.64 | 0.71 | 0.41 | 0.41 | 0.56 | 0.43 |
| Ti | 0.67 | 0.64 | -0.06 | 0.47 | 0.42 | 0.01 |

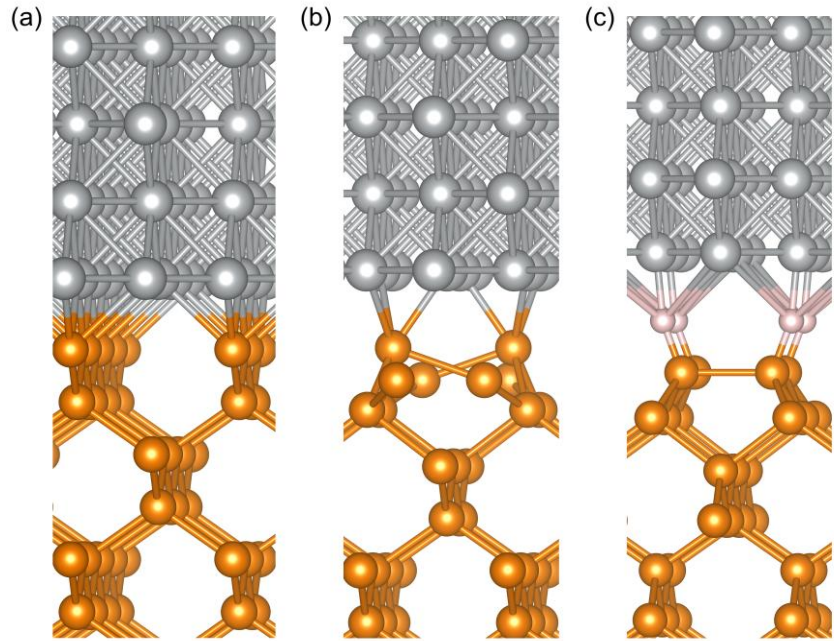

Fig. SM-2. Schematic atomic structures of Ag-Ge (001) interface with (a) $c(1\times1)$ non-reconstructed configuration, (b) $p(2\times2)$ dimer reconstruction configuration and (c) hydrogen-passivated configuration. The orange, pink and silver spheres represent Ge, H, and Ag atoms, respectively.

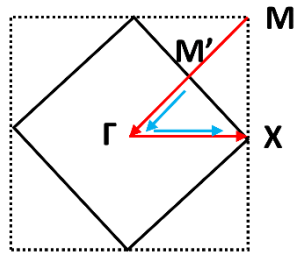

Fig. SM-3. The surface Brillouin zone and high symmetry points. Outer dashed line and red arrow represent $c(1\times1)$ unit cell and inner straight line and blue arrow represent $p(2\times2)$ unit cell.

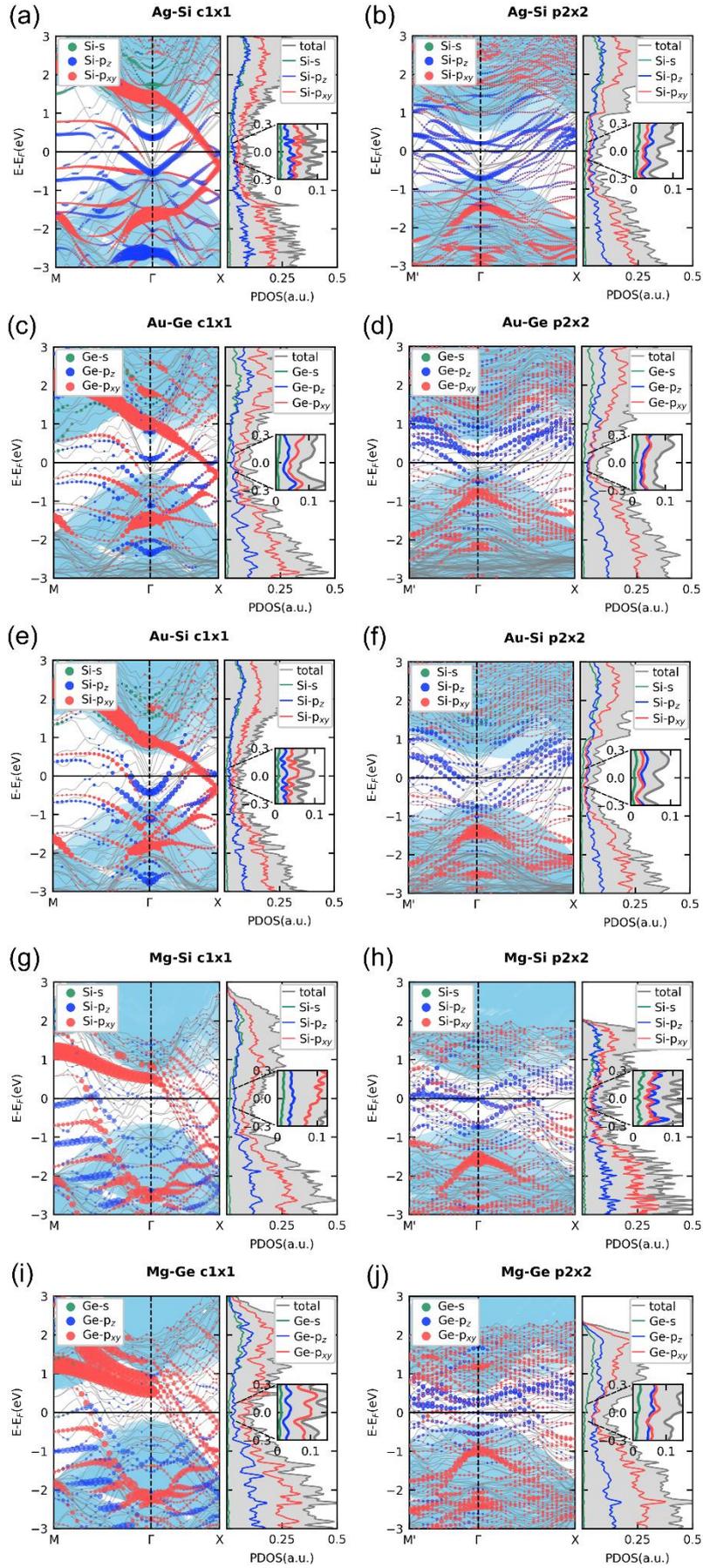

Fig. SM-4. The first-principles calculated interface band structure of different metal-Ge and metal-Si interface. The metal-Ge(Si) slab band structures are projected onto the first layer of Ge(Si) atoms and the density of states (PDOS) are projected onto all Si layers for (a) $c(1\times1)$ non-reconstructed interface configuration and (b) $p(2\times2)$ dimer reconstruction interface configuration, respectively. In the left panels, the light-blue shadow represents the bulk band structure projected to the (001) surface BZ; the gray lines denote the interface band structure, and the colored dots represent Ge(Si) atomic orbitals with dot size accounting for the relative amplitude. The insets zoom the PDOS near the Fermi level.

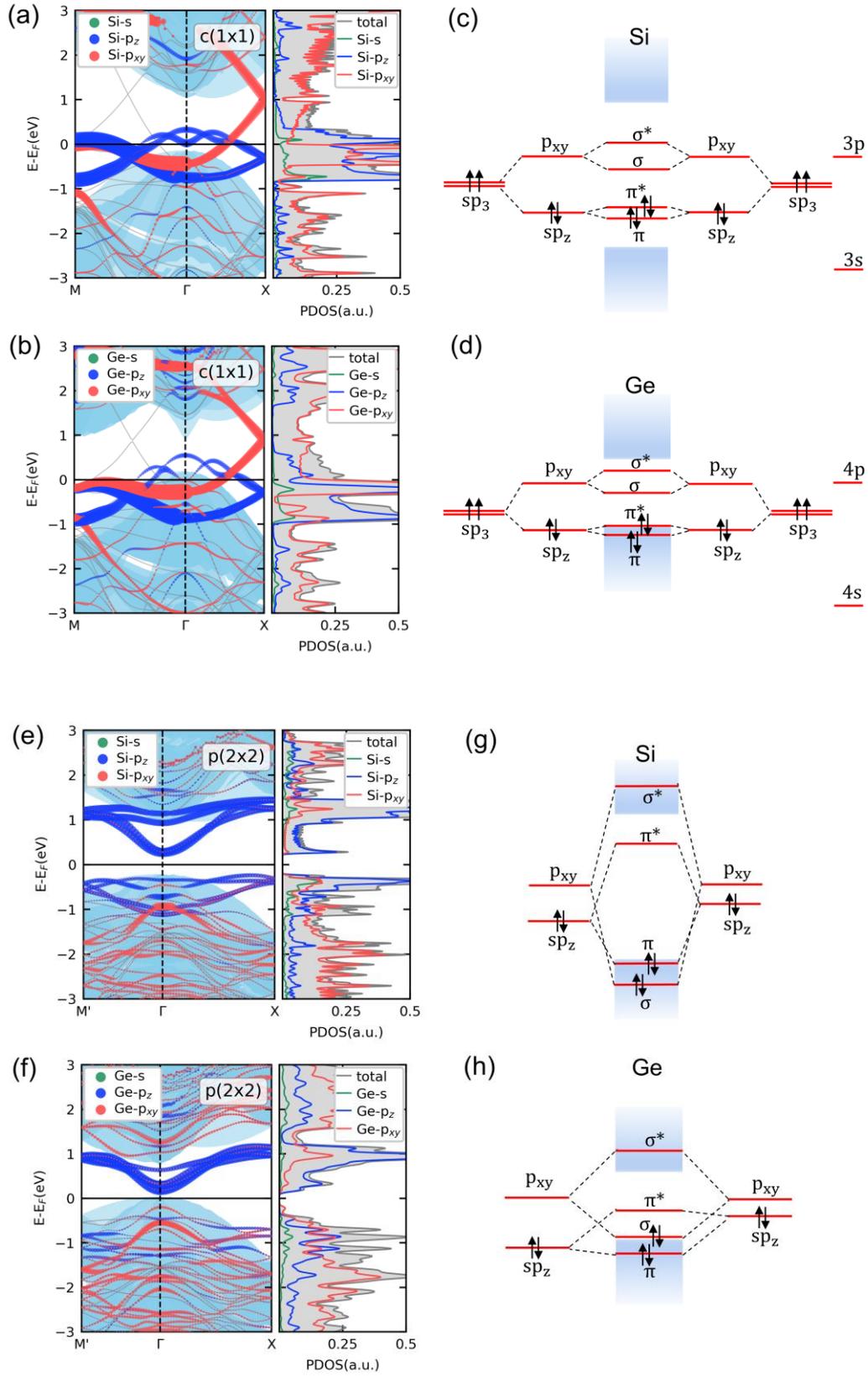

Fig. SM-5. The first-principles calculated surface band structure and the schematic diagram of orbital energy level shift. The band structure and density of states are projected onto the first atomic

layer of bare Si and Ge (001) surface with two different configurations: *c*(1×1) non-reconstructed configuration in (a) Si and (b) Ge with schematic diagrams illustrating the corresponding energy levels in (c) and (d), *p*(2×2) dimer reconstruction configuration in (e) Si and (f) Ge with schematic diagrams illustrating the corresponding energy levels in (g) and (h).

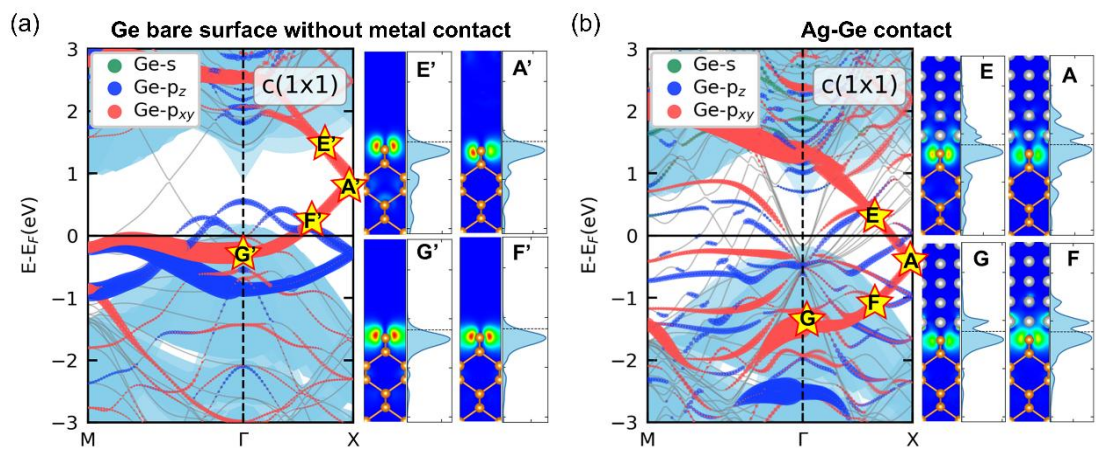

Fig. SM-6. Partial charge calculation of the selected states (a) in bare Ge (001) surface shown in Fig. SM-5(b), and (b) in Ag-Ge interface shown in Fig. 2(a). The orange spheres denote Ge atoms, and the silver spheres denote Ag atoms.

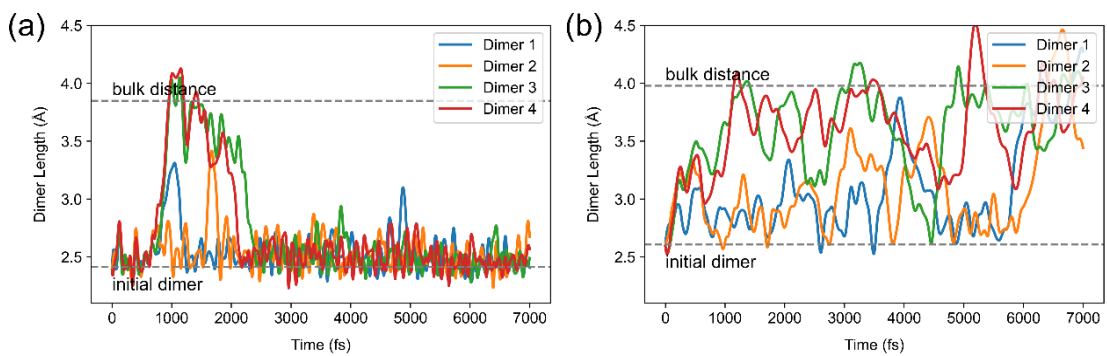

Fig. SM-7. Time evolution of the dimer bond lengths under molecular dynamics simulations at 300K. Simulations were performed with a supercell expansion to a p(4×2) surface structure, where each unit cell contains four dimers in both (a) Ag-Si (001) interface and (b) Ag-Ge (001) interface. The higher dashed lines indicate the equilibrium distances at 0K between adjacent surface atoms along the [110] direction in the non-reconstructed configuration, while the lower dashed lines represent

the equilibrium dimer bond lengths at 0K in the reconstructed configuration.

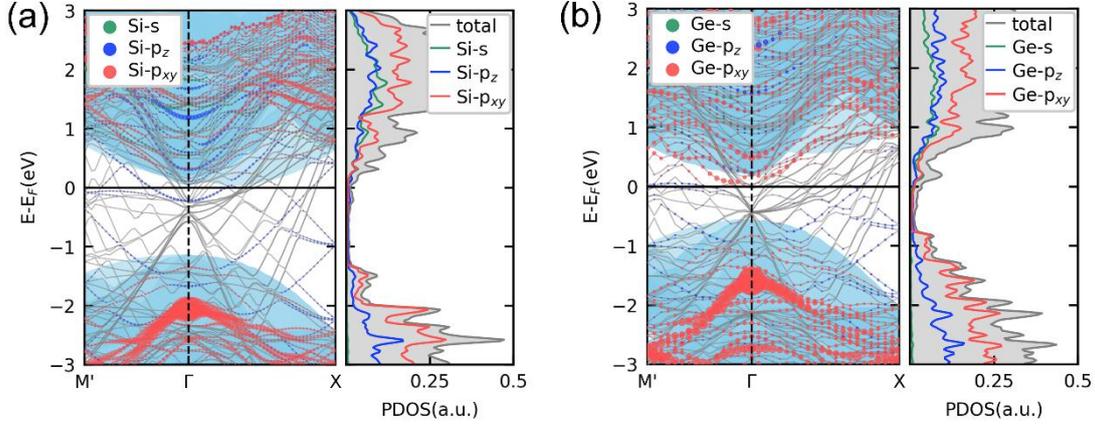

Fig. SM-8. The first-principles calculated interface band structure under hydrogen-passivation. (a) Ag-Si (001) interface and (b) Ag-Ge (001) interface, by projecting the electronic states onto the first layer of semiconductor and the projected density of states (PDOS) of all semiconductor layers.

**Metal-diamond (100) interface calculations**

The inset in Fig. SM-9 shows the optimized structures of diamond (001) surface in both non-reconstructed and reconstruction configurations. Experimental studies have shown that diamond (001) exhibit reconstruction mechanisms similar to those of Si and Ge [9]; however, diamond undergoes surface reconstruction with symmetric dimers to maintain its semiconducting surface properties, which differs from the asymmetric dimers observed in Si and Ge. Similarly, the band gap was corrected to match the experimental value (Eg=5.45eV) using the HSE functional with a mixing coefficient a=0.27. The structure modeling and computation procedures for the metal-diamond interface are the same as those for the metal-Si (001) and metal-Ge (001) interfaces described previously.

Considering computational costs and for simplicity, we only considered three typical metals (Ag, Rh, and Ir), which span a relatively wide range of metal work functions. The work functions were also recalculated to fit the strained metal, while the lateral lattice parameters of the diamond slabs were kept fixed. The results are summarized in Table SM-III and plotted in Fig. SM-9, where we obtained the pinning factor *S*. These results are consistent with previous theoretical studies, which also found that dimer-reconstructed diamond (001) exhibits S~0.29 [10]. Furthermore, the experimental pinning factor S~0.35 for diamond (001) [11-14] is also close to the value obtained from our theoretical dimer-reconstructed model. This supports the experimental tendency of diamond (001) surfaces to stabilize in the reconstructed configuration since the shorter dimer bond length of diamond (001) leads to stronger dimer bonds compared to those in Si [9].

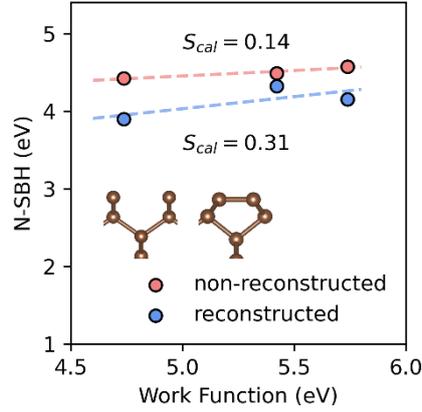

Fig. SM-9. Calculated n-type Schottky barrier height (N-SBH) of metal-diamond (001) interfaces under different configurations. The inset depicts the schematic atomic structures of reconstructed and reconstruction configurations, respectively.

Table SM-III. Calculated n-type Schottky barrier heights (N-SBH) data for metal-diamond (001) contacts in two different interface configurations. The $c(1\times1)$ represents the non-reconstructed structure, and $p(2\times2)$ represents the reconstructed structure. The calculated work functions (cal WFs) have been adjusted such that the lattice parameters are set equal to the optimized lattice constant of diamond ($a_C$ = 3.55 Å)

| Metal | cal WF (eV) | $c(1\times1)$: N-SBH (eV) | $p(2\times2)$: N-SBH (eV) |
|---|---|---|---|
| Ag | 4.74 | 4.42 | 3.90 |
| Rh | 5.42 | 4.49 | 4.32 |
| Ir | 5.74 | 4.57 | 4.15 |